\documentclass[%
reprint,
superscriptaddress,
onecolumn,
amsmath,amssymb,
aip,
]{revtex4-1}
\usepackage{color}
\usepackage{hyperref}
\usepackage{float}
\usepackage{graphicx}
\usepackage{textcomp}
\usepackage{epsfig}
\usepackage{epstopdf}
\usepackage{natbib}
\usepackage{multirow}
\usepackage{siunitx}
\usepackage{braket}
\usepackage[titletoc]{appendix}
\usepackage{dcolumn}
\newcolumntype{N}{D..{3.14}} 
\newcolumntype{Z}{D..{3.18}}

\usepackage{dcolumn}
\usepackage{bm}

\begin{document}
\preprint{APS/123-QED}

\title{High-concurrence time-bin entangled photon pairs from optimized Bragg-reflection waveguides}
\author{H. Chen}
\affiliation{Institut f\"{u}r Experimentalphysik, Universit\"{a}t Innsbruck, Technikerstra{\ss}e 25, 6020 Innsbruck, Austria}%
\affiliation{Department of Physics, National University of Defense Technology, Changsha, 410073, People's Republic of China}

   \author{S. Auchter}%
   \affiliation{Institut f\"{u}r Experimentalphysik, Universit\"{a}t Innsbruck, Technikerstra{\ss}e 25, 6020 Innsbruck, Austria}

   \author{M. Prilm\"uller}%
   \affiliation{Institut f\"{u}r Experimentalphysik, Universit\"{a}t Innsbruck, Technikerstra{\ss}e 25, 6020 Innsbruck, Austria}

   \author{A. Schlager}%
   \affiliation{Institut f\"{u}r Experimentalphysik, Universit\"{a}t Innsbruck, Technikerstra{\ss}e 25, 6020 Innsbruck, Austria}

   \author{T. Kauten}%
   \affiliation{Institut f\"{u}r Experimentalphysik, Universit\"{a}t Innsbruck, Technikerstra{\ss}e 25, 6020 Innsbruck, Austria}

   \author{K. Laiho}%
   \affiliation{Technische Universit\"at Berlin, Institut f\"ur Festk\"orperphysik, Hardenbergstr.~36, 10623 Berlin, Germany}

   \author{B. Pressl}%
   \affiliation{Institut f\"{u}r Experimentalphysik, Universit\"{a}t Innsbruck, Technikerstra{\ss}e 25, 6020 Innsbruck, Austria}

   \author{H. Suchomel}%
   \affiliation{Technische Physik, Universit\"at W\"urzburg, Am Hubland, 97074 W\"urzburg, Germany}

   \author{M. Kamp}%
   \affiliation{Technische Physik, Universit\"at W\"urzburg, Am Hubland, 97074 W\"urzburg, Germany}

   \author{S. H\"ofling}%
   \affiliation{Technische Physik, Universit\"at W\"urzburg, Am Hubland, 97074 W\"urzburg, Germany}
   \affiliation{School of Physics \& Astronomy, University of St Andrews, St Andrews KY16 9SS, UK}

   \author{C. Schneider}%
   \affiliation{Technische Physik, Universit\"at W\"urzburg, Am Hubland, 97074 W\"urzburg, Germany}

   \author{G. Weihs}%
   \affiliation{Institut f\"{u}r Experimentalphysik, Universit\"{a}t Innsbruck, Technikerstra{\ss}e 25, 6020 Innsbruck, Austria}

%

\begin{abstract}
Semiconductor Bragg-reflection waveguides are well-established sources of correlated photon pairs as well as promising candidates for building up integrated quantum optics devices.  Here, we use such a source with optimized non-linearity  for preparing time-bin entangled photons in the telecommunication wavelength range.  By taking advantage of pulsed state preparation and efficient free-running single-photon detection, we drive our source at low pump powers, which results in a strong photon-pair correlation. The tomographic reconstruction of the state's density matrix reveals that our source exhibits a high degree of entanglement. We extract  a concurrence of $88.9\pm 1.8\%$ and a fidelity of  $94.2 \pm 0.9\%$ with respect to a Bell state.
\end{abstract}


\date{\today}%

\maketitle

\section{Introduction}
Robust entangled photon sources are vital for performing quantum optics tasks fast and reliably. For example in a variety of quantum communication applications, no matter whether performed on the ground or via a satellite \cite{Rudolph-Why-optimistic-2016, Ma2012,Yin-Pan-Satellite-2017,Guenthner2017}, integrated quantum resources can be very useful. If the entangled photon sources that are often based on bulk crystals are replaced with integrated optics, one does not only greatly save space and reach better scalability but one also gains in optical stability\cite{brien--science--2007,politi--science--2008}.

Semiconductors provide an interesting integrated optics platform for preparing photon pairs via nonlinear optical effects, such as parametric down-conversion (PDC) \cite{lanco2006semiconductor,sarrafi2013continuous,Gregor-Monolithic-Source-2012}. By utilizing  Bragg-reflection waveguides (BRWs) made of Al$_{x}$Ga$_{1-x}$As with $x$ being the aluminum concentration, we can benefit from  their large effective second-order optical nonlinearity \cite{Boyd2003} and the flexibility in their design \cite{Pressl-2017-advanced-BRW}. Moreover, for constructing large-scale fiber-optic networks, these structures  have a broad transparency window in the telecommunication wavelengths and they benefit from the electro-optic capability of AlGaAs, so that both active and passive optical elements can be realized\cite{adachi1993properties, Gehrsitz-refractive-index-2000, Dietrich2016}. Well-established fabrication technologies are available, which render these monolithic structures altogether excellent for nonlinear integrated optics.

In order for the PDC process to occur, the interacting light modes need to fulfill energy and momentum conservation  \cite{West--2006--Analysis-BRW,Helmy--2006--Phase-matching, helmy2011recent}. Recently, BRWs based on the interaction of fundamental and higher order spatial modes have become popular for generating collinearly propagating pairs of photons -- usually called signal and idler -- \cite{Gregor-Monolithic-Source-2012,Gunthner-2015,Claire2016Multi-user,kang2016monolithic} and a lot of effort has been put into optimizing their performance. For example, BRWs with a very low birefringence have proven to be versatile sources of polarization entangled states \cite{Valles-2013,Horn-scientific-reports-2013,schlager--2017--temporally,kang2015two}. Additionally, such semiconductor waveguides  are good candidates for producing photon pairs with electrical pumping, in other words, by integrating the pump laser with the PDC source on the same chip \cite{Boitier-Ducci-Electrically-2013,bijlani2013semiconductor}.

Semiconductor waveguides have also been shown to emit time-energy entangled photons \cite{Sarrafi2014, Autebert2016} that were first treated by Franson  in 1989 \cite{Franson-PhysRevLett-1989}. However, in order to achieve a high degree of entanglement these realizations require a highly coherent pump laser. A more versatile variant, using a pulsed excitation scheme, is  time-bin entanglement \cite{brendel1999pulsed}, in which a photon pair is created into a coherent superposition of two subsequent time bins with a well-defined relative phase. The pulsed operation is convenient for performing quantum optics tasks fast, and when being transmitted over long distances in optical fibers, time-bin entangled photons are more robust against decoherence  than the polarization-entangled ones due to the inevitable polarization-mode dispersion \cite{dynes-2009-efficient-entanglement}.

In the past, these types of entanglement have been demonstrated on various quantum photonic platforms ranging from PDC in bulk crystals \cite{marcikic-2004-entanglement-distribution, Kwon-2013-Time-bin-spdc-cw, vedovato-genuine-time-bin} and waveguides \cite{tanzilli2002ppln, Ma2009}, and  spontaneous four wave mixing in silicon waveguides and optical fibers \cite{takesue2014entangled-time-bin,takesue-2005-time-bin-optical-fiber} to photon emission from quantum dots \cite{Gregor-time-bin-dot-2014,prilmuller-2017-hyper-entanglement}. These sources, nevertheless, work only in highly controlled environments, suffer from low conversion and background suppression efficiencies, or they cannot be efficiently miniaturized and integrated with active components, like lasers.

Here, we demonstrate time-bin entanglement from a BRW sample that has been optimized to possess a high optical nonlinearity \cite{Pressl-2017-advanced-BRW}.
By combining pulsed pumping with efficient and fast free-running photo-detection, our PDC emitter produces a strong photon-pair correlation \cite{Covi2015} and a low level of noise.
Thereafter, we employ state tomography to reconstruct the density matrix of the time-bin entangled state. We demonstrate that BRWs can meet the demand of producing these states with a high degree of entanglement.
Finally, with the help of the achieved concurrence we can demonstrate the quality of the entangled state and predict that a violation of the Bell's inequality is possible with our source.

\section{BRW sample and photon-pair characteristics}

In our experiments we utilize a multicore BRW sample as described in Ref.~\cite{Pressl-2017-advanced-BRW}.  The \SI{365}{\nano\meter} thick BRW core of Al$_{0.428}$Ga$_{0.572}$As is surrounded by \SI{398}{\nano\meter} thick inner matching layers of Al$_{0.2}$Ga$_{0.8}$As below and above. Around this core region are \SI{356}{\nano\meter} thick outer matching layers of Al$_{0.628}$Ga$_{0.372}$As and 6/5 layers of Al$_{0.2}$Ga$_{0.8}$As/Al$_{0.628}$Ga$_{0.372}$As having thicknesses of \SI{127}{\nano\meter}/\SI{443}{\nano\meter}. The sample is designed in a way that the largest aluminum concentration can be achieved by summing up the two lower concentrations, which simplifies the wafer growth process via molecular beam epitaxy. The used BRW sample is reactive-ion plasma etched to just above the core, has a length of \SI{2}{\milli\meter} and a ridge width of \SI{4}{\micro\meter}. Its degeneracy wavelength of \SI{1534}{\nano\meter} was determined via second harmonic generation. This type-II process occurs between cross-polarized total-internal reflection modes in the telecom range and the Bragg modes in the near infrared, which are higher order spatial modes.

\begin{figure}
\centering
\includegraphics[width = 0.8\textwidth]{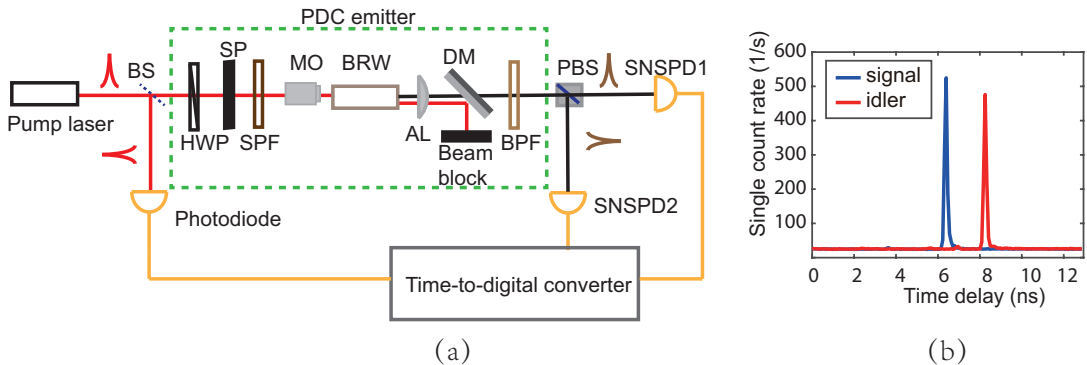}%
\caption{\label{fig:waveguide_schem} (a) Measurement setup for investigating the properties of photon-pair emission from BRW. (b) Histograms of detection events in SNSPD1 and SNSPD2 with respect to the time elapsed from the detection of the laser trigger at the pump power of \SI{60}{\micro\watt}. The two curves are shifted for better visibility. Abbreviations:  AL: aspheric lens, BPF: band-pass filter,  BS: beam sampler, DM: dichroic mirror,  HWP: half-wave plate, SPF: short pass filter, MO: microscope objective, PBS: polarizing beam splitter, SNSPD: superconducting nanowire single-photon detector, SP: sheet polarizer. }
\end{figure}

First, we explore the performance of our BRW by investigating the characteristics of the emitted photons. For this purpose, we use the experiment described in \autoref{fig:waveguide_schem}(a). A pulsed Ti:Sapphire laser (\SI{76.2}{\mega\hertz} repetition rate, \SI{767}{\nano\meter} central wavelength and \SI{0.8}{\nano\meter} bandwidth) is used as a pump for the PDC process.  A small fraction of the pump beam is sent to a fast photodiode via a beam sampler to generate an electronic trigger signal for synchronizing the measurement devices. Thereafter, the beam passes through a polarization control set up of a half-wave plate and sheet polarizer and through a short pass filter. The pump beam is then coupled into the BRW with a 100$\times$ microscope objective, whereas the light coupled out of the BRW is collimated with an aspheric lens. A dichroic mirror separates the residual pump beam from the PDC emission, which is then sent through a  spectral filter that has a \SI{12}{\nano\meter} bandwidth and is centered at the degeneracy wavelength. Filtering is used to limit the spectral range of the PDC emission and to suppress residual background illumination. A polarizing beam splitter separates the cross-polarized signal and idler beams, which are collected with aspheric lenses into single mode fibers and detected with superconducting nanowire single-photon detectors (SNSPDs) (SingleQuantum Eos) that are optimized for the telecom C-Band and have detection efficiencies better than 60\%. In our setup, their combined rates of measured background light and dark counts are \SI{360\pm 60}{/\second} 
for the signal channel and  \SI{390\pm 80}{/\second} 
for the idler channel. Finally, a time-to-digital converter is used to record time stamps of the detected photons and the laser trigger.

We start by investigating the single counts in signal and idler, shown in \autoref{fig:waveguide_schem}(b) for an average incident pump power  of approximately \SI{60}{\micro\watt} measured in front of the BRW before the microscope objective.  Due to the pulsed pump it is justified to apply time gatings  of \SI{0.5}{\nano\second} at the detection. We achieve gated single count rates of  \SI{1210(40)}{/\second} and \SI{1090(40)}{/ \second}
for signal and idler, respectively. The difference between the measured single counts in signal and idler is caused by the slightly different coupling efficiencies into the single mode fibers connected to the two SNSPDs. Additionally, we extract a signal-to-noise ratio as high as 20.8(9) in the single counts that verifies the low level of total remaining background illumination. Moreover, the achieved coincidence count rate of \SI{46(7)}{/\second} is almost perfectly free from spurious counts.

\begin{figure}[!b]
\includegraphics[width = 0.4\textwidth]{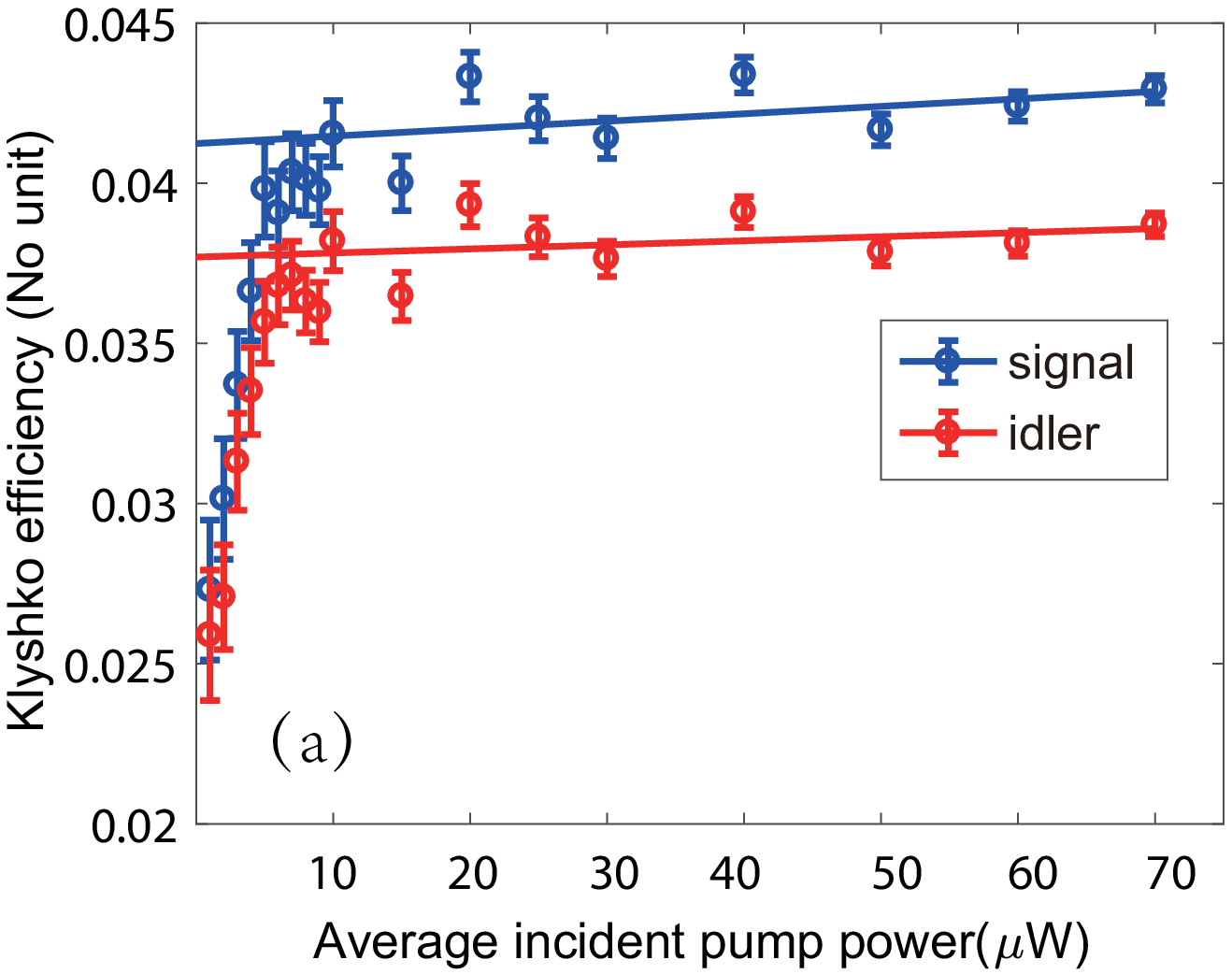}
\hspace{1ex}
\includegraphics[width = 0.4\textwidth]{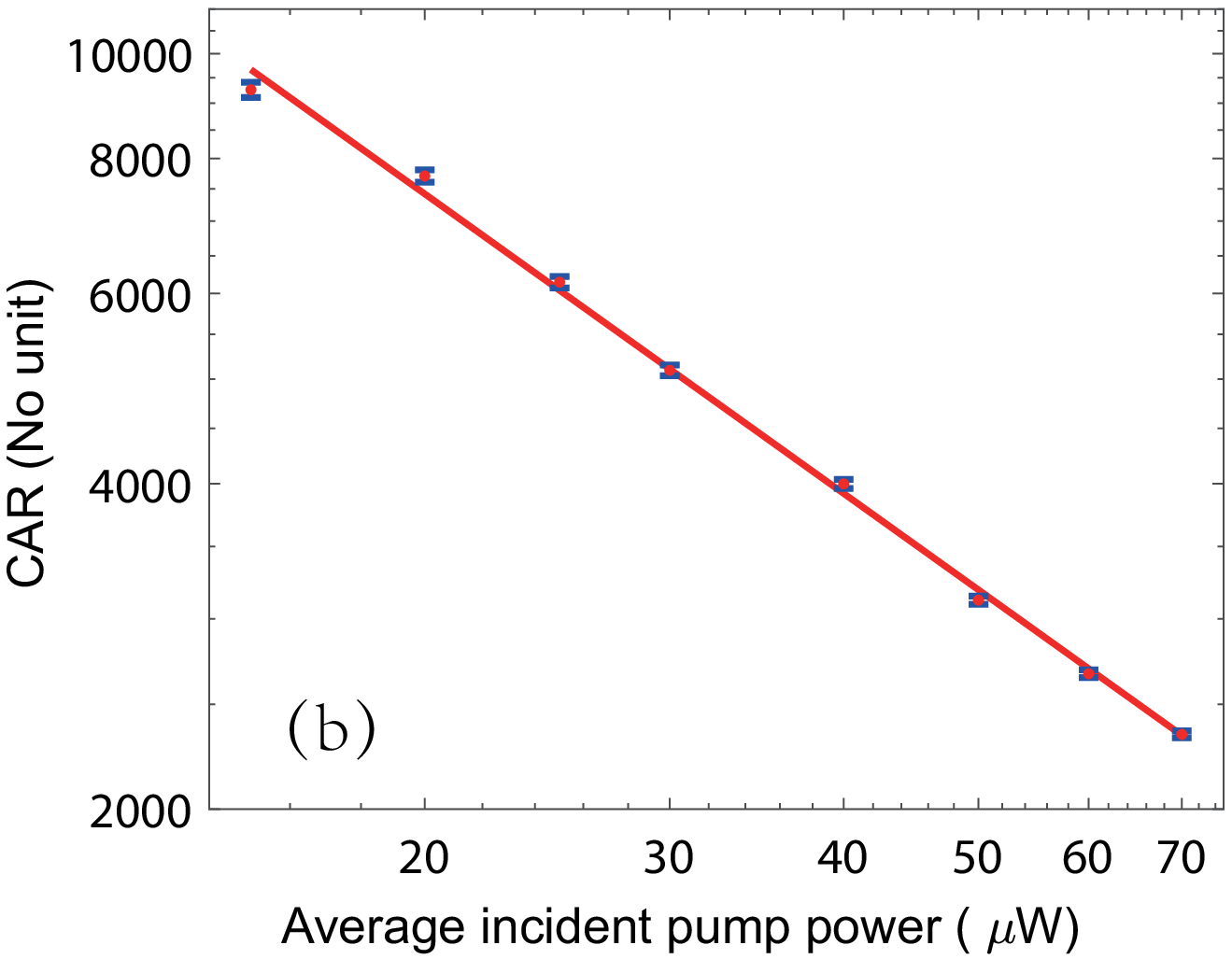}
\caption{\label{fig:Klyshko_CAR} (a) Klyshko efficiency  for signal and idler and (b) CAR with respect to the average incident pump power. Symbols indicate the measured values, whereas the solid lines are fitted.}
\end{figure}
\begin{figure}
\centering
\end{figure}

Next, we measure a power series of singles and coincidence counts to characterize the photon-pair properties of the emitted PDC light. We evaluate the Klyshko efficiency~\cite{Klyshko1977}  for signal (idler)  as the ratio of coincidence counts to the single counts in idler (signal) with respect to the pump power as shown in  \autoref{fig:Klyshko_CAR}(a).
We see that at incident pump powers below about \SI{10}{\micro\watt} this ratio is limited by the background noise, which contributes to the gated single counts and lowers the Klyshko efficiency. Additionally, the Klyshko efficiency is expected to grow with increasing pump power due to the higher photon numbers in the PDC emission. Therefore, we exclude the apparent drop of the Klyshko efficiency at weak pumping from our linear fits,  extrapolate them to low pump powers and obtain the values of  \SI{4.12\pm 0.09}{\%} and \SI{3.77\pm0.08}{\%} for the total collection efficiencies of signal and idler, respectively. Moreover,  we emphasize that this measurement  sets a  lower bound for the usable pump power to gain well-defined PDC emission.

The power series of coincidence counts shows a linear behavior as expected for a highly multimodal PDC process and delivers the source brightness. The coincidence counts increase with a rate of \SI{750 \pm 30}{counts/\second/\milli\watt} 
with respect to the average incident pump power measured before the microscope objective.
This figure of merit describes how well the pump light can be converted to pairs of photons that are finally detected.
Thus, it is affected by all experimental light coupling and detection imperfections as well as the strength of the nonlinearity and the waveguide losses. It is also affected by the confinement of modes and their interaction length.
In our experiment,  the rather high losses of the waveguide  and  the poor coupling of the pump light into the Bragg-mode each decrease the BRWs' brightness by about an order of magnitude \cite{Pressl--2015}. When corrected for these effects, the achieved brightness of our BRW compares well to the conventional sources \cite{M.Fiorentino2007}.

Finally, we explore the coincidences-to-accidentals ratio (CAR), which is a loss-independent measure of the photon-pair correlation between signal and idler. We estimate the CAR via $R_{C}/R_{A}$, in which $R_{C}$ is the measured rate of coincidences and the accidental rate is estimated via $R_{A} = R_{s}R_{i}/R_{t}$, where $R_{\mu}$ ($\mu = s,i$) is the measured rates of singles in signal ($s$) and idler ($i$) and $R_{t}$ is the trigger rate. In \autoref{fig:Klyshko_CAR}(b)  we present the CAR for our BRW with respect to the average incident pump power in the region where the PDC emission dominates over any background [see \autoref{fig:Klyshko_CAR}(a)] and achieve a CAR as high as $9260 \pm 150$.
Investigating the CAR on a logarithmic scale reveals, whether the  power dependencies of the single  and coincidence counts deviate from each other. If both are strictly linear, we expect to find a slope of $-1$. Instead we find \SI{-0.918\pm 0.015}{}, which indicates minor imperfections in the photon-pair process and its detection. Nevertheless, the achieved values indicate a strong photon-number correlation between signal and idler.

\section{Preparation and characterization of time-bin entangled states}

\begin{figure*}
\centering
\includegraphics[width = 0.95\textwidth]{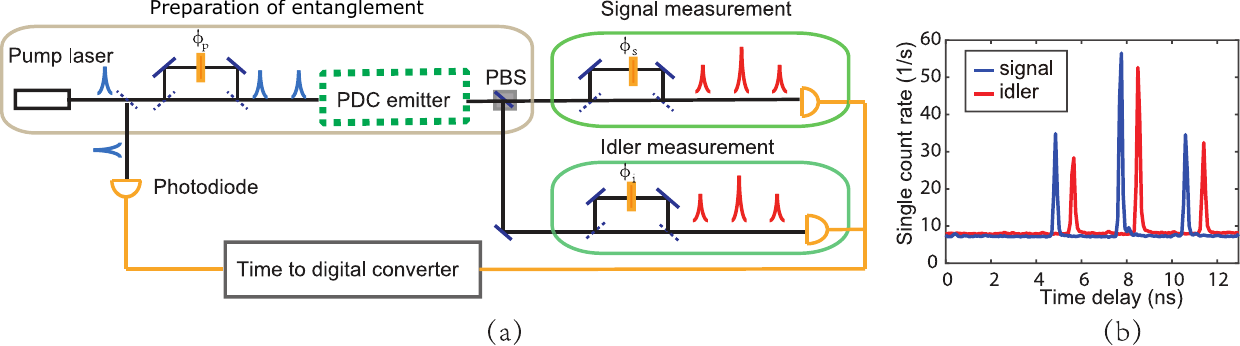}
\caption{\label{fig:time-bin-diagram} (a) Measurement setup for generating and detecting time-bin entanglement using the PDC emitter from \autoref{fig:waveguide_schem}. (b) Histograms of the detection events in SNSPD1 and SNSPD2 with respect to the time elapsed from the detection of the laser trigger at an incident pump power of \SI{100}{\micro\watt}. The unbalanced Mach-Zehnder interferometers are used for the entanglement generation and detection. The phase of each interferometer is controlled by rotating glass plates on motorized stages placed in the interferometers long arms.}
\end{figure*}

In order to prepare and detect time-bin entangled states we extend our setup in \autoref{fig:waveguide_schem}(a) and now use the setup depicted in \autoref{fig:time-bin-diagram}(a). Before coupling the pump laser with its incident power of \SI{100}{\micro\watt} to our BRW, we employ an imbalanced Mach-Zehnder  interferometer to transform the pulsed pump into a coherent superposition of two pulses that define the so-called ‘early’ and ’late’ time bins. The time delay between these pulses is about \SI{3}{\nano\second}, which is much shorter than the time of \SI{13.1}{\nano\second} between two successive pump laser pulses.  Additionally, before detection, we send signal and idler through similar interferometers having the same delay between the ‘early’ and ‘late’ time bins as the pump interferometer. In the experiment, the three interferometers are combined in a single multi-path free-space setup similar to the one reported in Ref.~\cite{Gregor-time-bin-dot-2014}. For measuring the interference fringe patterns we integrate over \SI{2}{\min} at each setting, while we increase this integration time to \SI{20}{\min} for doing the state tomography in order to grow the ensemble size and statistical accuracy. Active stabilization with a reference beam helps compensating long term phase drifts in all three distinct spatial modes.

For achieving a maximally entangled state, we require that the 'early' and 'late' pump pulses generate photon pairs with the same probability. Additionally, the probability to generate two photon pairs, one in the 'early' and one in the 'late' pump pulse, has to be negligible. If these conditions are met, the photon pairs are emitted in the time-bin entangled state

\begin{equation}
\ket{\Phi} = \frac{1}{\sqrt{2}}\big(\ket{0}_s\ket{0}_i + e^{i\phi_p}\ket{1}_s\ket{1}_i\big),
\label{eq:phi}
\end{equation}
in which $\ket{0}_{\mu}$ and $\ket{1}_{\mu}$  denote the early and late states and the variable $\phi_p$ is the pump interferometer's phase.

\autoref{fig:time-bin-diagram}(b) illustrates histograms of the detected single count events for both signal and idler. We can clearly see three distinct peaks. As expected, the central peaks are nearly twice as high as the surrounding ones. The leftmost peak corresponds to the events, where the signal and idler photons are created by the early pump pulse and they both pass through the short paths of the detection interferometers. Similarly, in the case of the rightmost events, the pump and PDC photons traverse the long paths of the interferometers. Since photon pairs produced from an 'early-pump/long-analysis-path' are indistinguishable from the 'late-pump/short-analysis-path' case, the cross-correlation of the central peaks of signal and idler shows interference fringes, depending on the relative phase of the output interferometers.

\begin{figure}
\centering
\includegraphics[width = 0.4\textwidth]{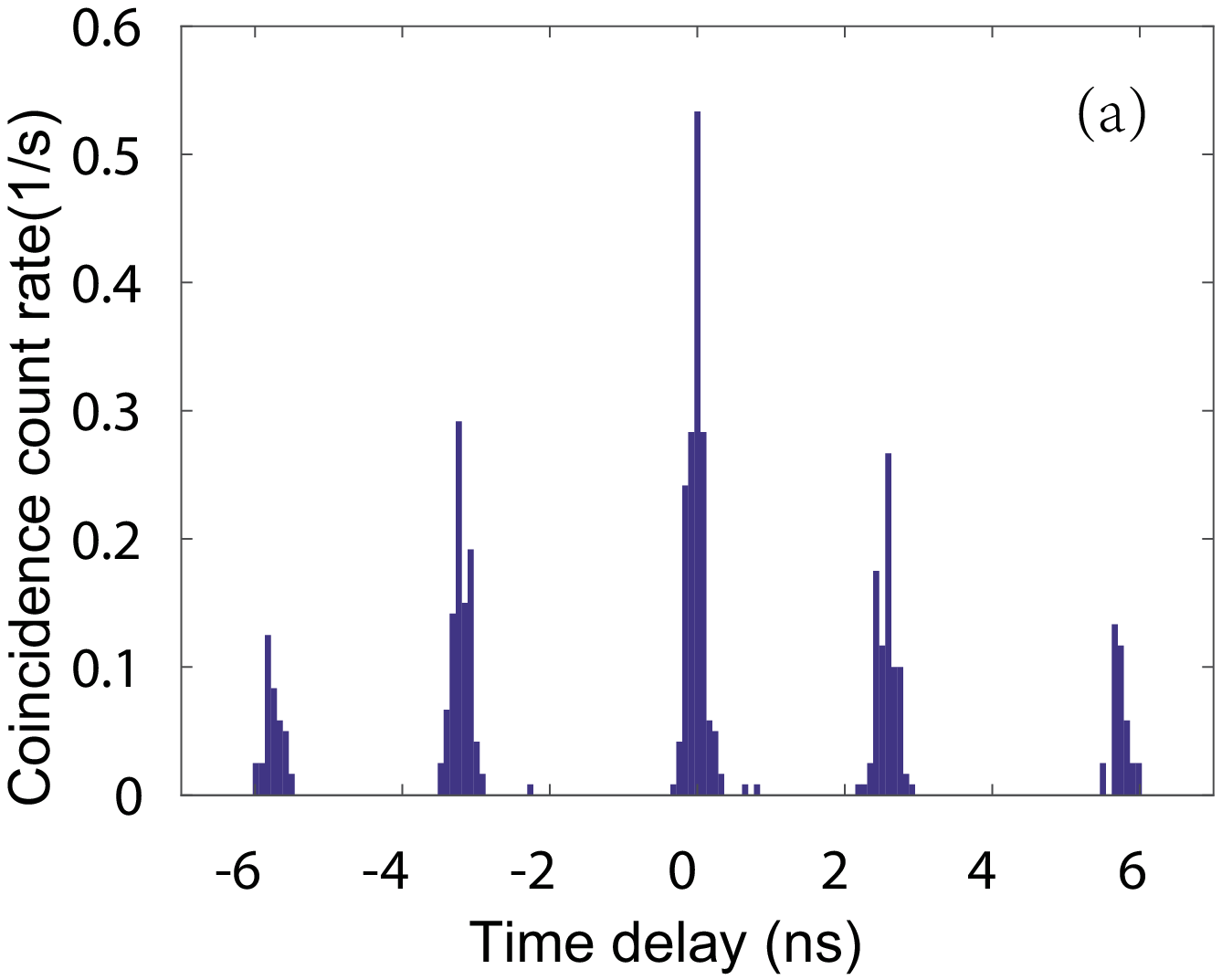}
\hspace{1ex}
\includegraphics[width = 0.4\textwidth]{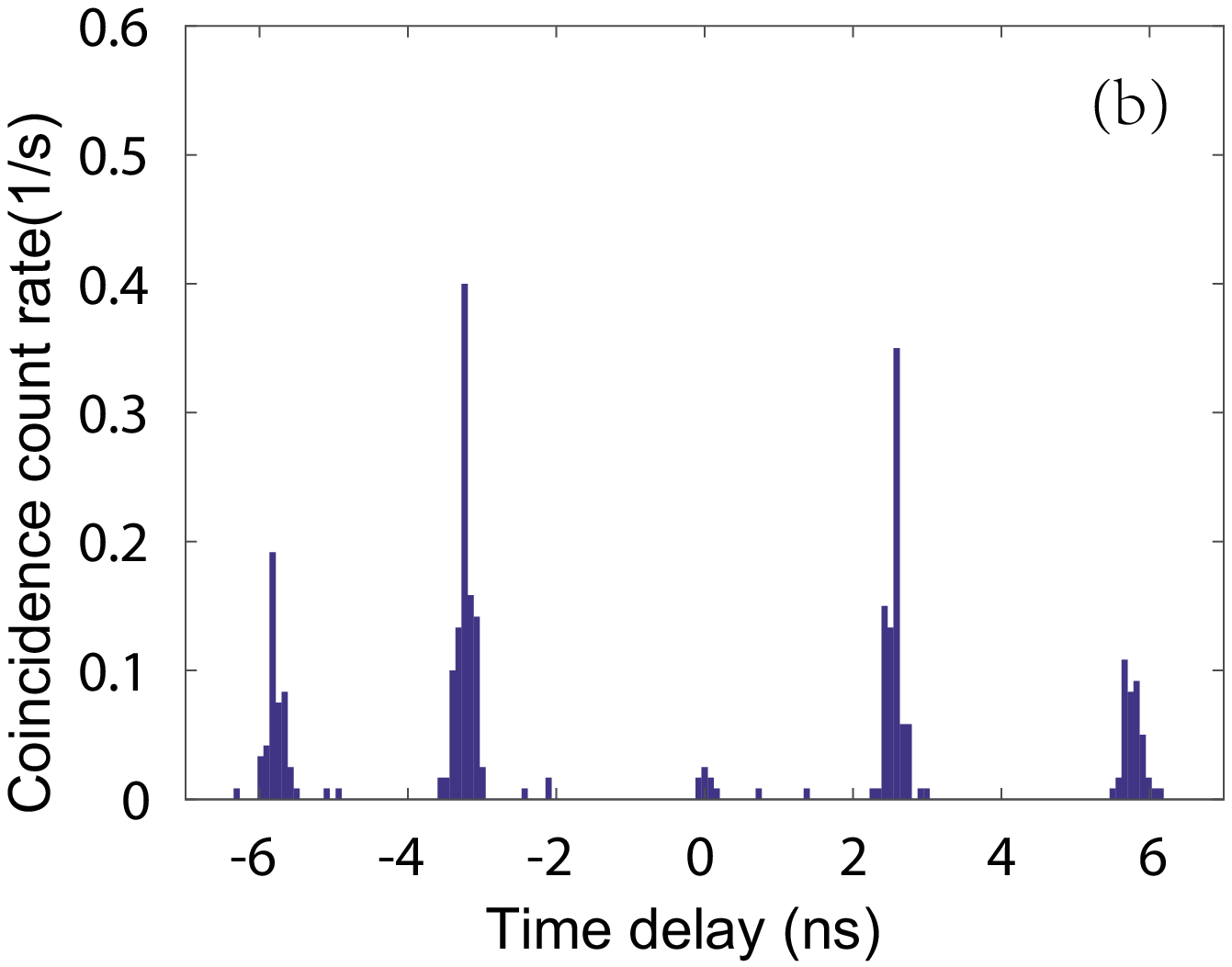}
\caption{\label{fig:time-bin-peaks} Interferograms of coincidences between signal and idler shown in \autoref{fig:time-bin-diagram}(b) with respect to the time delay between their detection at two relative signal and idler interferomenters' phases  resulting in (a) maximal and (b) minimal amount of coincidences.}
\end{figure}

In  \autoref{fig:time-bin-peaks} we illustrate this interference by presenting the signal and idler coincidences within the five possible discrete time delays between their detections for two cases that result in the maximal and minimal amounts of measured coincidences, respectively.  Again we employ time gates of \SI{0.5}{\nano\second} width to separate the coincidence peaks from each other.
The rate of coincidences  $R$ in the central peak oscillates  sinusoidally with  respect to the phase change  in the pump, signal and idler interferometers and is given by \cite{Marcikic2002}
\begin{equation}
R \propto 1 - V\cos(\phi_s + \phi_i - \phi_p),
\end{equation}
in which $\phi_s$ and $\phi_i$ are the phases of the signal and idler interferometer, respectively, and  $V$ denotes the fringe visibility.
 While the visibility in the time basis is typically almost perfect, we emphasize that $V$ is a measure of the quality of the  interference when the state is projected into a superposition of the time bins. Therefore, it is often degraded due to the state's impurity.

\begin{figure}[H]
\centering
\includegraphics[width = 0.4\textwidth]{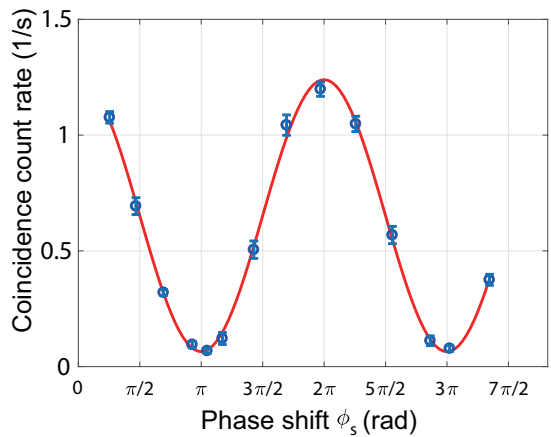}
\caption{\label{fig:visibility} Measured visibility (symbols) with respect to the phase change in the signal interferometer. The solid line represents a sinusoidal fit. }
\end{figure}

\autoref{fig:visibility} shows the measured coincidence count rate in the central time bin with respect to the phase  in the signal interferometer including the data in \autoref{fig:time-bin-peaks}. This phase can be changed by rotating the glass plates placed in the interferometers long arms. We  observe a fringe visibility of $90.2 \pm 0.9\%$ 
 without any compensation of accidental counts or subtraction of other background contributions.  We subject the slight degradation in the visibility to minor imperfections in the mode overlap of the used  interferometer. The effect of spurious counts on the visibility can be estimated via the CAR, which was in the measurement 530(50). 
If just the finite CAR was responsible, the highest achievable visibility is  99.62(4), which is well-above our experimental value.

To fully characterize the entangled state prepared in \autoref{fig:visibility}, we perform a tomographic reconstruction of the density matrix by projecting the prepared state onto four different bases that we define as $\ket{0}$, $\ket{1}$, $\ket{+X}=\frac{\ket{0}+\ket{1}}{2}$ and $\ket{+Y} = \frac{\ket{0}+i\ket{1}}{\sqrt{2}}$.
By performing 16 correlation measurements between all combinations of \{$\ket{0}$,$\ket{1}$,$\ket{+X}$,$\ket{+Y}$\}, we reconstruct the $4\times 4$ density  matrix of our time-bin entangled state.
For this purpose, we only need four different measurements with the signal and idler interferometers' phases ($\phi_s$, $\phi_i$) set to the values of (\SI{0}{\degree}, \SI{0}{\degree}), (\SI{0}{\degree}, \SI{90}{\degree}) and (\SI{90}{\degree}, \SI{90}{\degree}), (\SI{90}{\degree}, \SI{0}{\degree}) that reveal the visibilities in the superposition bases with  \SI{0}{\degree} corresponding to $\ket{+X}$ and   \SI{90}{\degree}  to $\ket{+Y}$, respectively \cite{takesue-2009-tomography}.
For this purpose,  the glass plates' angles are calibrated via the fringe pattern measurement. We assign the phase (\SI{0}{\degree},\SI{0}{\degree}) to angle settings at which a maximum amount  of coincidences are expected in \autoref{fig:visibility}. By finding the coincidence minimum that corresponds to a phase change to (\SI{0}{\degree},\SI{180}{\degree}) we can extract the angle settings required for the state tomography.

Finally, we reconstruct the density matrix using the maximum likelihood method described in Ref.~\cite{banaszek1999maximum, Kwiat2005StateTomo}. The errors are recovered via Monte-Carlo simulation. The absolute values (abs) of the elements of the reconstructed density matrix $\rho$ are given by
%

\begin{figure}
\begin{minipage}{0.4\linewidth}
\centering
\includegraphics[width = 1\textwidth]{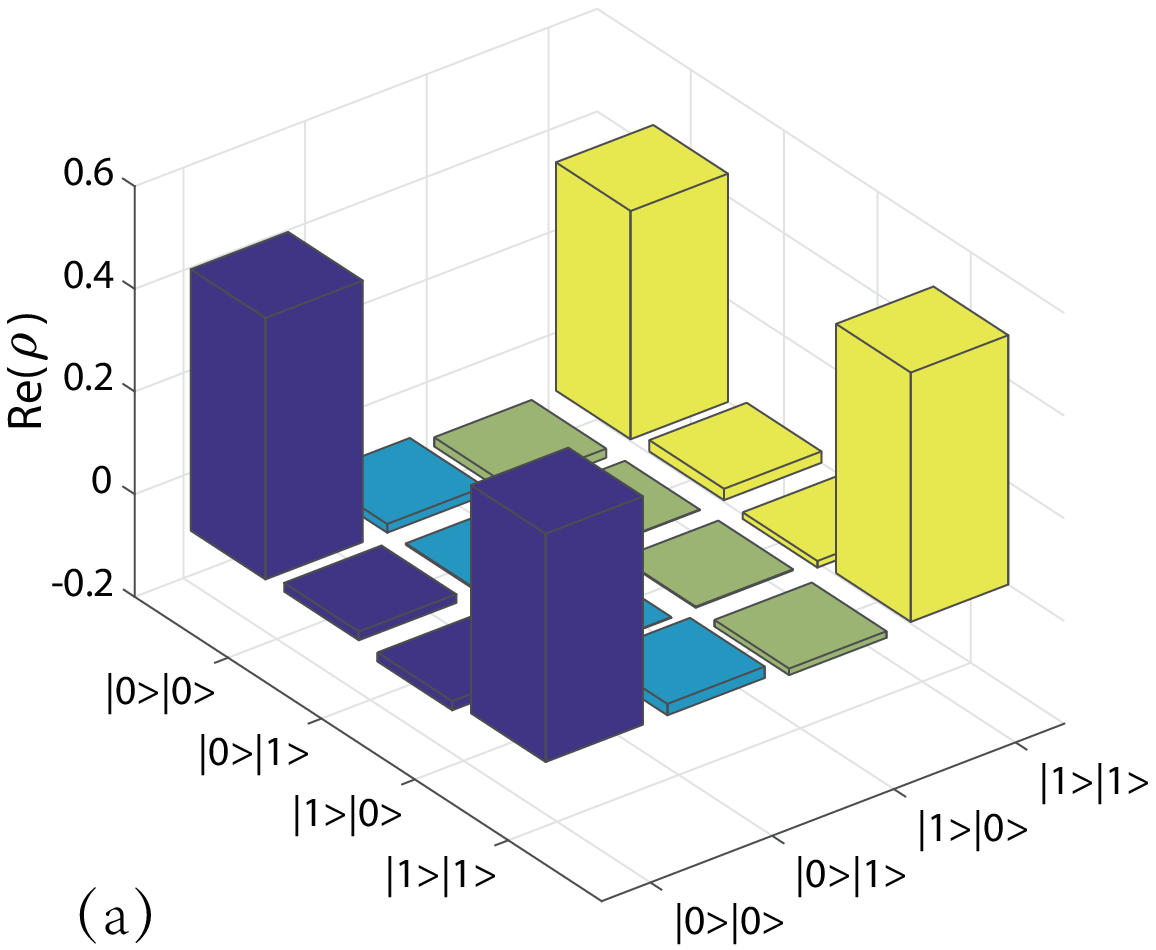}
\end{minipage}
\begin{minipage}{0.4\linewidth}
\centering
\includegraphics[width = 1\textwidth]{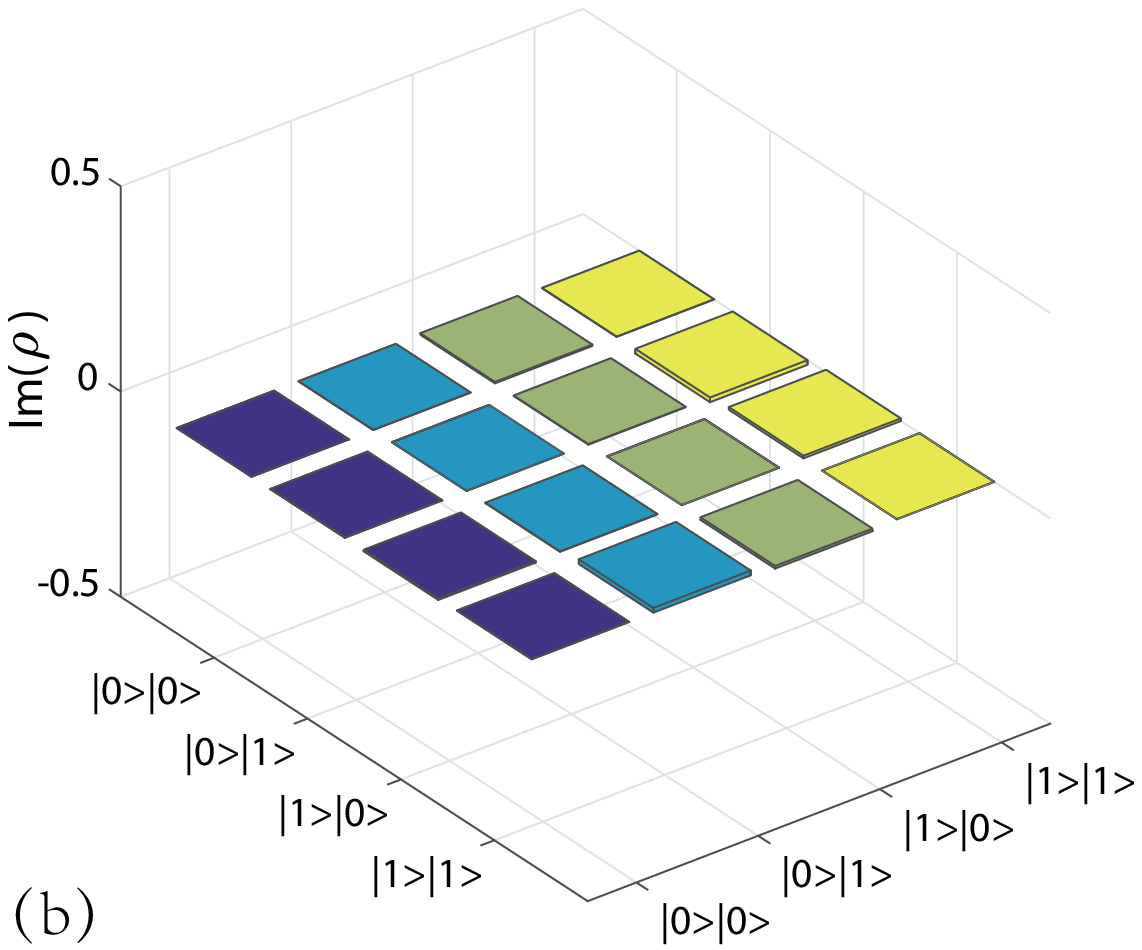}
\end{minipage}
\caption{\label{fig:tomography} (a) Real and (b) imaginary parts of the reconstructed density matrix given in Eq.~(\ref{eq:DM}).}
\end{figure}

\arraycolsep=-30pt

\begin{eqnarray}
\label{eq:DM}
\textrm{abs}(\rho) =
\frac{1}{100}\cdot
\left[ \hspace{8ex}
\begin{array}{ZZZZ}
50.9(8) & 1.7(7) & 1.8(7) & 44.5(9)\\
1.7(7) & 0.3(1) & 0.17(7) & 2.5(7)\\
1.8(7) & 0.17(7) & 0.21(9) & 1.4(7)\\
44.5(9) & 2.5(7) &  1.4(7) & 48.6(8)
\end{array}
\hspace{-8ex} \right], \hspace{3ex}
\end{eqnarray}
%
and its  real and imaginary parts are shown in \autoref{fig:tomography}. The obtained density matrix includes as expected four main contributions and only very small undesired background. The weights of the diagonal elements reflect the division of the power in the pump beam interferometer and the main off-diagonal elements are mainly real valued as also expected from \autoref{fig:visibility}.

From the density matrix in \autoref{fig:tomography}  we obtain a concurrence of $88.9\pm 1.8\%$ and a fidelity of $94.1\pm 0.5\%$ with respect to the state $\ket{\Phi^+} = \frac{1}{\sqrt{2}}(\ket{0}_s\ket{0}_i + \ket{1}_s\ket{1}_i)$. As the former quantifies the high degree of the entanglement achieved, the latter describes the probability of producing the target state $\ket{\Phi^+} $.  Moreover, the obtained concurrence is high enough to violate Bell's inequality. Following Ref.~ \cite{Verstraete2002} we can estimate boundaries for the Bell-parameter $\mathcal{S}$ formulated by Clauser-Horne-Shimony-Holt (CHSH) via the concurrence and find that $\mathcal{|S|}= 2.51(5)...2.68(5) > 2$ for  the prepared time-bin state  highlighting the suitability of  BRWs as low noise PDC emitters for entangled state generation. 

\section{Conclusion}
Integrated optics devices provide means for miniaturizing the bulk optics resources that are still today largely used in many quantum optics tasks. BRWs based on semiconductor materials are indeed good candidates to become truly practical  as integrated quantum photonic components. We used a BRW sample with a simplified epitaxial structure designed for having a larger nonlinearity and to ease up the fabrication process for producing cross-polarized signal and idler beams via PDC. By utilizing efficient, free-running detectors at the single photon level, we can drive our source at low pump powers enabling to achieve a modest Klyshko efficiency and a strong photon-pair correlation. Additionally, the power dependency of the CAR indicates only a small number of detected spurious counts. We prepared time-bin entangled states from the PDC emission of our BRW and achieved a high degree of entanglement without subtracting any spurious counts. We obtained a concurrence of $88.9\pm 1.8\%$ that is high enough to violate the Bell's inequality and a fidelity of $94.2\pm 0.9\%$ to the $\ket{\Phi ^+}$ state.
Our results offer means to develop BRWs that reach the low level background noise required for the preparation of high quality entangled states from photon pairs.  Additionally, we believe that further integration and miniaturization of our PDC emitter and the bulky interferometers will result in higher brightness and in better overlap of the modes required for achieving a higher concurrence. Altogether, our investigations pave the way of utilizing BRWs as integrated sources of entangled photon pairs in the telecommunication wavelength range in pulsed operation with high repetition rates.

\section{Acknowledgments}
This work was supported by the Austrian Science Fund (FWF) through the project nos. I-2065-N27 and J-4125-N27, the DFG project no. SCHN1376/2-1, the ERC project \textit{EnSeNa} (257531), the State of Bavaria and China Scholarship Council (201503170272). We thank  A. Wolf and S. Kuhn for assistance during sample growth and fabrication and J. Loitzl for laboratory assistance.

\providecommand{\noopsort}[1]{}\providecommand{\singleletter}[1]{#1}%

\end{document}